# Unconditional Stability Analysis of *N*-Port Networks Based on Structured Singular Value Computation

Aimar Mateo
Dpto. Electricidad y Electrónica
University of the Basque Country (UPV/EHU)
Leioa, 48940, Spain,
aimar.mateo@ehu.eus

Ibone Lizarraga
Dpto. Electricidad y Electrónica
University of the Basque Country (UPV/EHU)
Leioa, 48940, Spain
ibone.lizarraga@ehu.eus

Jorge Terrer
Dpto. Electricidad y Electrónica
University of the Basque Country (UPV/EHU)
Leioa, 48940, Spain
jorge.terrer@ehu.eus

Aitziber Anakabe
Dpto. Electricidad y Electrónica
University of the Basque Country (UPV/EHU)
Leioa, 48940, Spain
aitziber.anakabe@ehu.eus

J.M. Collantes
Dpto. Electricidad y Electrónica
University of the Basque Country (UPV/EHU)
Leioa, 48940, Spain
juanmari.collantes@ehu.eus

***Abstract*—In this paper, a novel approach based on robust stability concepts and tools is introduced to evaluate the unconditional stability of microwave active *n*-port devices. An efficient calculation of the Structured Singular Values of the *n*×*n* scattering matrix is proposed to obtain the stability characteristics of the device. The presented method is validated in two ways. First, it is applied to a referential 4x4 scattering parameter set for independent verification. Second, the method is applied to a 4-port GaAs FET amplifier fabricated in hybrid technology. The results confirm the validity and computational efficiency of the proposed approach.**

## I. Introduction

Microwave amplifier designers are acquainted with the undesired oscillations that may show up when the load terminations at the device ports are varied. Small-signal stability analyses based on well-known analytical parameters ($K$ or $\mu$ factors) are widely used to verify the unconditional stability of 2-port devices versus load variations in the microwave domain [1], [2]. The situation is more complex if the analysis has to be performed at more than 2-ports. This may be the case if we want to include the effect of the load impedances at the bias lines of the amplifier or if multiport amplifier structures are studied.

For the case of a 3-port device, analytical solutions for unconditional stability verification still exist [3] although they are less straightforward and intuitive than the 2-port case. For $n$-port devices with $n \geq 4$ there is no analytical solution for the definition of unconditional stability factors. The most thorough and advanced works on unconditional stability analysis, both theoretical and practical, for $n$-port circuits with $n \geq 4$, are presented by Colangeli et al. [4], [5]. Global optimization methods based on method of moments are suggested in [4], [5] to solve a non-convex optimization problem. The computational cost of these approaches limits their practical application to $n \leq 4$ ports.

In this work, robust stability concepts are used to evaluate the unconditional stability of microwave active $n$-port devices. Specifically, we propose the calculation of the Structured Singular Values of the $n$x$n$ scattering matrix of the active circuit. Although there is no closed-form method to obtain the Structured Singular Value of a matrix of order $n \geq 4$, there exist efficient techniques to compute its upper and lower bounds [6], [7]. When these limits are very similar, they give an accurate answer to the evaluation of unconditional stability.

The method is validated in two different ways. In the one hand, we use the artificial 4×4 scattering parameter set generated in [4] as external reference for independent verification. On the other hand, the procedure is applied to a 4-port amplifier fabricated in hybrid technology, with two ports dedicated to the RF input and output and the other two located at the bias lines.

## II. Structured Singular Values

Robust stability theory for control systems is deeply studied in control theory to guarantee that an intrinsically stable system is not de-stabilized by any possible feedback perturbation bounded inside some admissible limits [8]. Robust stability concepts can be applied to the analysis of the unconditional stability of multiport active microwave networks versus load variations at their ports.

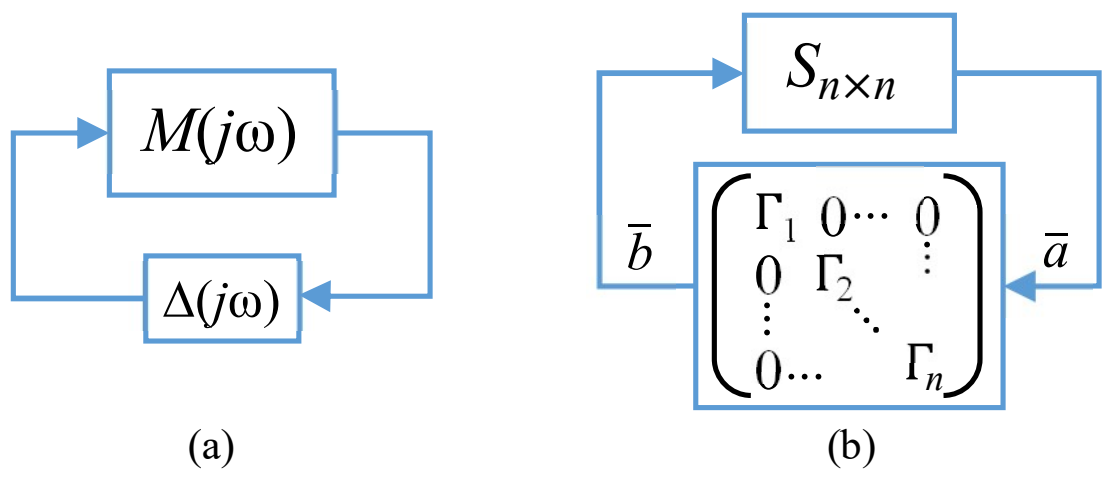

Fig. 1. (a) Generic $M$-$\Delta$ feedback system. (b) Feedback representation of an $n$-port linear network loaded with $n$ terminations, where $\bar{a}$ and $\bar{b}$ are the incident and reflected power wave vectors.

Consider the $M$-$\Delta$ feedback loop of Fig. 1(a). $M(j\omega)$ is a frequency dependent matrix representing a linear MIMO (Multiple Input – Multiple Output) internally stable system. $\Delta(j\omega)$ is the frequency dependent matrix associated to the feedback network. The $M$-$\Delta$ feedback system is robustly stable if there is no $\Delta(j\omega)$ that makes $(I - M\Delta)$ singular at some frequency $\omega$. A sufficient condition for robust stability is given by the small gain theorem [8]:

$$\|M(j\omega)\|_\infty \cdot \|\Delta(j\omega)\|_\infty < 1 \quad (1)$$

where $\|\cdot\|_\infty$ stands for the $H_\infty$-norm of a linear system. Note that the $H_\infty$-norm of a dynamic linear system is the supreme of its singular values and represents the maximum vector amplification of the system. If the $H_\infty$-norm of the feedback network $\Delta(j\omega)$ is bounded, such that:

$$\|\Delta(j\omega)\|_\infty \leq {}^{1}/_{\gamma} \quad (2)$$

with $\gamma$ being a real positive value, then, robust stability is guaranteed if the $H_\infty$-norm of system $M(j\omega)$ satisfies:

$$\|M(j\omega)\|_\infty < \gamma. \quad (3)$$

Inequality (3) is a sufficient condition for robust stability versus any kind of feedback $\Delta(j\omega)$. However, if $\Delta(j\omega)$ has a particular structure, like diagonal or block diagonal, the condition in (3) is far too conservative [8], [9]. A better parameter to characterize robust stability when $\Delta(j\omega)$ is diagonal is given by the Structured Singular Values (SSV) of the frequency dependent matrix $M(j\omega)$. Let us call simply $M$ the system matrix $M(j\omega)$ at a particular frequency $\omega$. The Structured Singular Value of the complex matrix $M$, $\mu(M)$, is defined as the inverse of the minimum of the $\Delta$'s maximum singular values, $\sigma_{max}$, that make $(I - M\Delta)$ singular [8], [9]:

$$\mu(M) \triangleq \frac{1}{min\{\sigma_{max}(\Delta)\ |\det(I-M\Delta)=0\}},\ \Delta \in \overline{\Delta} \quad (4)$$

where, in this case, $\overline{\Delta}$ refers to the class of matrices with a specific diagonal structure. Directly from the definition of $\mu(M)$, the $M$-$\Delta$ system is stable for all diagonal and bounded (2) $\Delta(j\omega)$, if and only if the SSVs of $M(j\omega)$ are $< \gamma$ at all frequencies:

$$\mu\big(M(j\omega)\big) < \gamma \quad \text{for all } \omega. \quad (5)$$

Let us particularize the above theory to a linear $n$-port microwave active network loaded with $n$ passive terminations at its ports (Fig. 1(b)). The linear system is described by an $n{\times}n$ matrix of scattering parameters $(M(j\omega) \equiv S(j\omega))$. The feedback $\Delta(j\omega)$ is now a diagonal matrix of reflection coefficients at the network's ports, $\Gamma_i$ with $i = 1$ to $n$, as in [5], [10]. The above robust stability concepts can be directly applied to evaluate the unconditional stability of the $n$-port active network $S(j\omega)$. For unconditional stability, load terminations inside the passivity limit are only considered, $|\Gamma_i| \leq 1$ for all $n$, which implies that $\gamma = 1$ in (2), (3) and (5). Therefore, unconditional stability of the active network $S(j\omega)$ is guaranteed if the SSVs of $S(j\omega)$ are $< 1$ at all frequencies:

$$\mu\big(S(j\omega)\big) < 1 \quad \text{for all } \omega. \quad (6)$$

Unfortunately, there is no closed-form method to compute the SSV of a matrix of order $n \geq 4$. However, there exist well-stablished algorithms to calculate upper and lower bounds for the SSV of a matrix [6], [7]. They are efficiently implemented in the MATLAB toolbox "Robust control toolbox", through the function *mussv*. The upper bound determination can be formulated as a convex optimization problem at each frequency, thus it can be easily computed. Note that this upper bound is tight for three or less ports [9]. When the upper and lower bounds agree, we have an accurate calculation of the SSV. When not, the upper bound gives at least a sufficient condition to evaluate unconditional stability.

Robust stability can also be expressed in terms of the stability margin introduced by Safonov [11] that is the reciprocal of the SSV. A parameter equivalent to Safonov's stability margin, the stability radius, was independently proposed by Colangeli et al. [5] for microwave $n$-port networks. In this work, we propose the use of the SSV directly to evaluate unconditional stability due to its efficient calculation using the *mussv* function.

## III. Validation

The proposed method based on computing the upper and lower bounds of the SSV has been tested in two separate 4-port case scenarios. First, we utilize an artificial S-parameter set generated in [4] for verification and benchmarking. This enables validation against an independent and external reference. Second, for a more realistic example, the method is validated on a medium power 4-port GaAs FET amplifier.

### A. *Generated* 4×4 *S-parameter set*

We use here the 4×4 S-parameter set that was generated in [4] (Table II, on page 4181) to allow the reader an independent validation of the results. To simulate a frequency variation, the $S_{11}$ parameter was scaled by a coefficient $c$ in [4]. To compare the results with those in [4], the SSV is evaluated for 1000 points of $c$, ranging from 0 to 3. The upper and lower bounds obtained with the function *mussv* are plotted versus $c$ in Fig. 2. Note that the two bounds are equivalent, thus, the calculation of SSV is accurate (except for a small deviation for values of $c > 2.6$ which is far from the stability bifurcation and does not contribute to the stability information). The system is potentially unstable for $c > 2.03$ ( $\mu(S) \geq 1$) in perfect agreement with the results in [4].

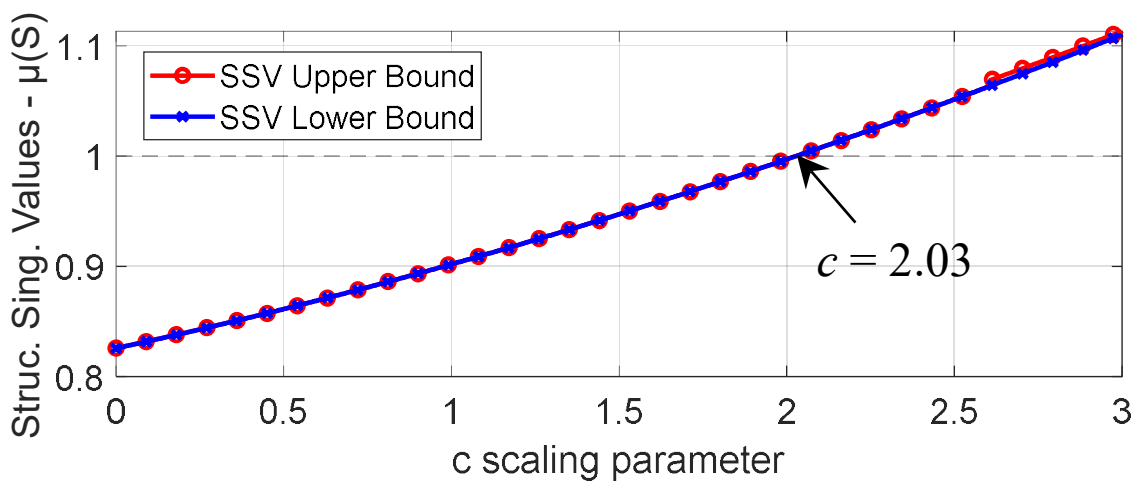

Fig. 2. Upper and lower bounds of the SSV versus the scaling parameter $c$, obtained from the generated 4x4 scattering parameter set in [4]. System is potentially unstable for $c > 2.03$.

The computation time to obtain the results of Fig. 2, with a grid of 1000 points, is only 2.2 s, demonstrating the efficiency of the SSV calculation. The calculation is performed in a computer equipped with an Intel Core i5-10500 CPU running at a clock speed of 3.10 GHz, paired with 32 GB of RAM.

### B. *4-port GaAs FET amplifier*

To further verify the reliability of the proposed method, the unconditional stability of a medium-power 4-port amplifier prototype built in microstrip technology (Fig. 3) has been analyzed. The amplifier is based on a FLU17XM GaAs FET transistor and operates at 1.2 GHz. Two of the ports are the input and output ports for the RF signal. The other two ports (G and D) are located at the bias lines to allow a characterization of the low-frequency resonances, as in [12].

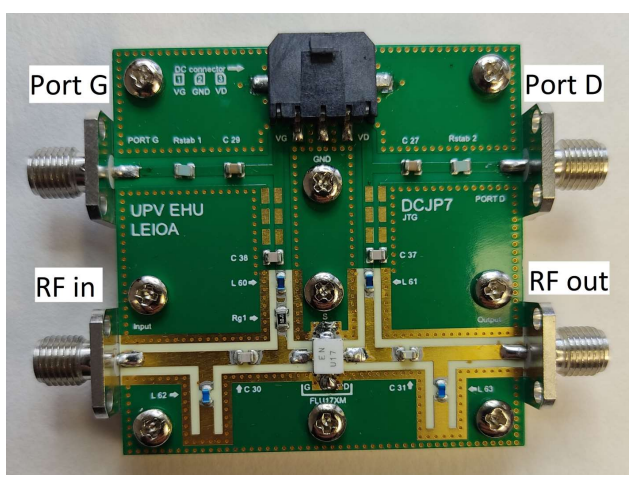

Fig. 3. 4-port amplifier based on a FLU17XM GaAs FET transistor.

The simulated 4×4 S-parameter matrix corresponding to the bias point $V_{GS}$ = -2 V, $V_{DS}$ = 2.1 V is considered first. For this bias, the amplifier is intrinsically stable when loaded with 50 Ω. This has been verified using pole-zero stability analysis [13] to check the Rollet proviso. The SSV upper and lower bounds are calculated up to 6 GHz, with a grid of 1000 frequency points (Fig. 4). The two bounds overlap, providing an accurate value for the SSV. Simulation time to obtain these curves is 1.8 s in the same computer. The results reveal potential instability within the frequency range of 703 MHz to 1.1 GHz, as the value obtained is greater than one, indicating the possible occurrence of oscillations under specific load conditions at the four ports.

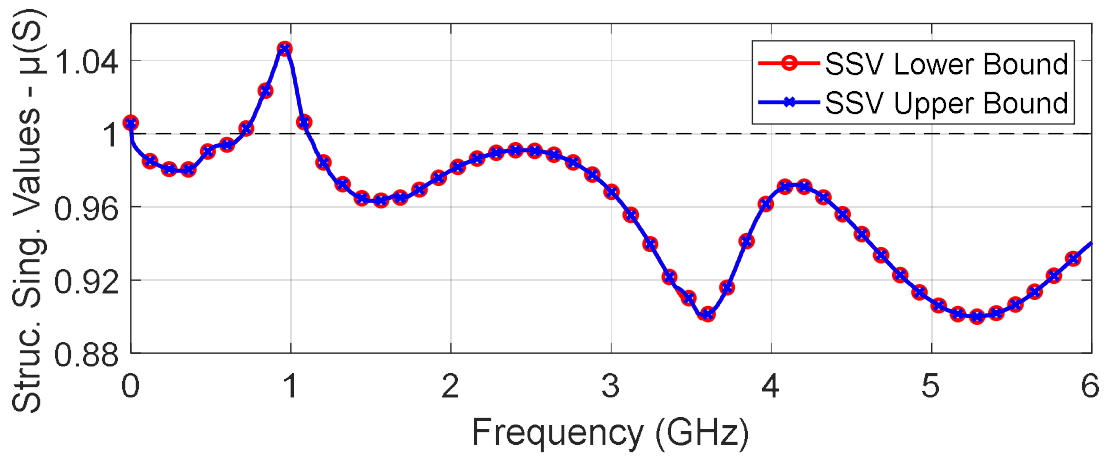

Fig. 4. Upper and lower bounds of the SSV obtained from the simulated S-parameter matrix for the bias point $V_{GS}$ = -2 V and $V_{DS}$ = 2.1 V.

This instability has been confirmed through measurements. A coupler is connected at the output port to monitor the output signal in a Spectrum Analyzer (SA). Lossless reflective loads implemented with delayed shorts are varied at the four ports to look for instabilities, (Fig. 5). An oscillation is experimentally found in the spectrum analyzer for a particular load configuration at the four ports (Fig. 6). The oscillation frequency is about 900 MHz, which falls within the previously identified range shown in Fig. 4.

The efficiency in computing the SSV through the *mussv* function facilitates conducting parametric stability analyses. As an example, Fig. 7 shows the calculated SSV obtained for $V_{DS}$ = 2.1 V and different values of $V_{GS}$, ranging from -1.9 V to -2.3 V. In all the cases the upper and lower bounds overlap, providing an accurate value for $\mu(S(j\omega))$. We can observe that the amplifier is unconditionally stable for $V_{GS} < -2.1$ V. Consistently, using the setup of Fig. 5, no load configuration has been found to produce oscillation in the circuit for $V_{GS}$ = -2.2 V.

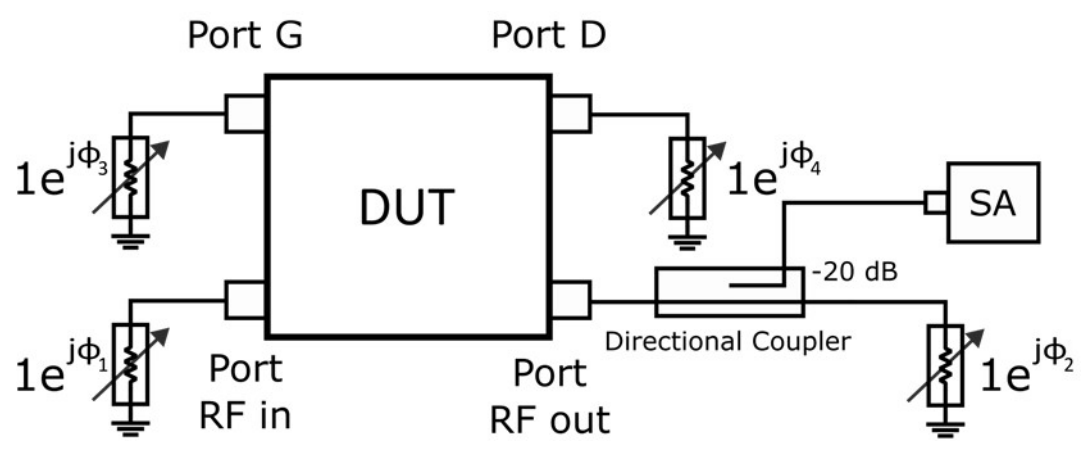

Fig. 5. Measurement set-up for the 4-port amplifier. Highly reflective loads are varied at the four ports to look for possible oscillations.

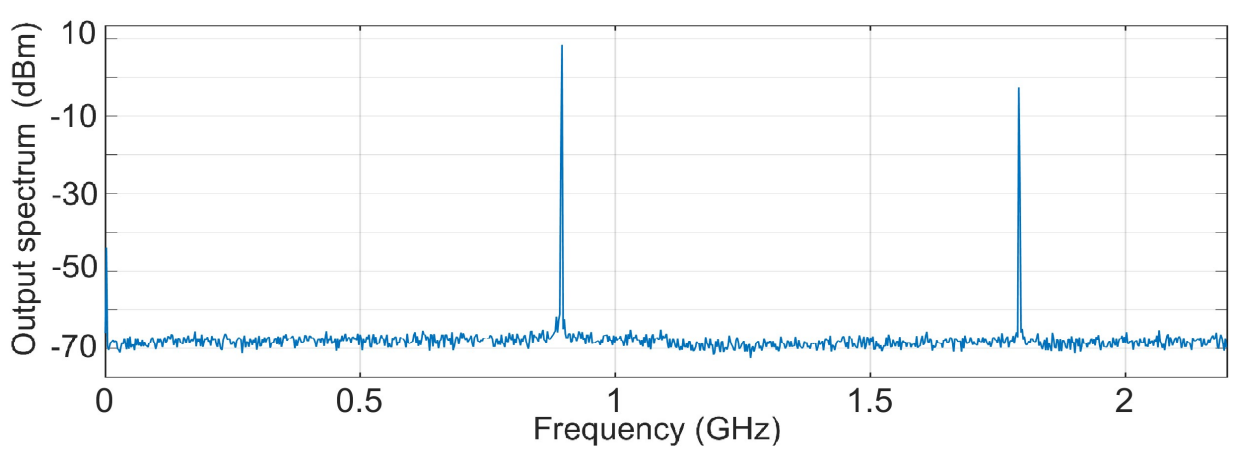

Fig. 6. Measured output spectrum for a particular set of reflective loads connected to the four ports. Oscillation around 900 MHz is measured for $V_{GS}$ = -2 V and $V_{DS}$ = 2.1 V.

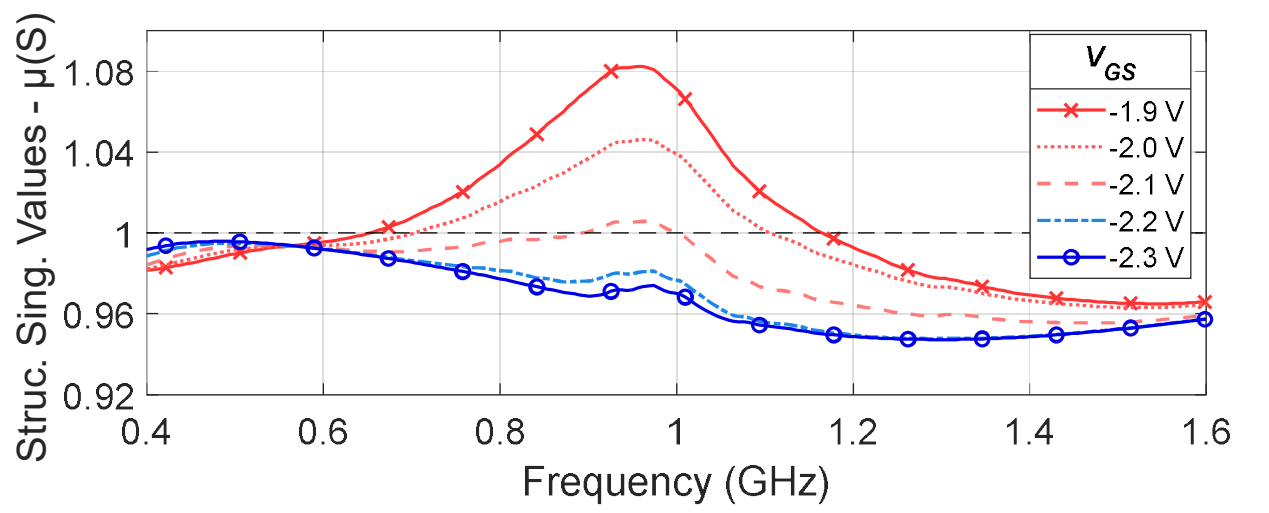

Fig. 7. Calculated SSV, for different values of $V_{GS}$ ranging from -1.9 V to -2.3 V with a fixed drain voltage of $V_{DS}$ = 2.1 V. Upper and lower bounds overlap for each $V_{GS}$ of the sweep. Zoom around 1 GHz.

## IV. Conclusions

In this paper, the application of the Singular Structured Value from robust control theory to evaluate the unconditional stability of microwave active *n*-port devices has been presented and validated. Independent validation has been performed using a set of scattering parameters generated by other researches in previous works. As a practical example, a second validation is carried out on a manufactured 4-port amplifier. Results have shown that the proposed methodology accurately predicts unconditional stability in complex multiport active devices while maintaining an efficient computational cost.